\documentclass[11pt,twoside]{article}
\usepackage{./asp2014}

\aspSuppressVolSlug
\resetcounters

\begin{document}

\title{M31N 2008-12a --- The Remarkable Recurrent Nova in M31}
\author{M.~J.~Darnley$^1$
\affil{$^1$Astrophysics Research Institute, Liverpool John Moores University, IC2 Liverpool Science Park, Liverpool, L3 5RF, UK; \email{M.J.Darnley@ljmu.ac.uk}}}

\paperauthor{M.J. Darnley}{M.J.Darnley@ljmu.ac.uk}{0000-0003-0156-3377}{Liverpool John Moores University}{Astrophysics Research Institute}{Liverpool}{Merseyside}{L3 5RF}{UK}

\begin{abstract}
M31N\,2008-12a is a remarkable recurrent nova within the Andromeda Galaxy.  With eleven eruptions now identified, including eight in the past eight years, the system exhibits a recurrence period of one year, and possibly just six months.  This short inter eruption period is driven by the combination of a high mass white dwarf ($1.38\,\mathrm{M}_\odot$) and high mass accretion rate ($\sim1.6\times10^{-7}\,\mathrm{M}_\odot\,\mathrm{yr}^{-1}$).  Such a high accretion rate appears to be provided by the stellar wind of a red giant companion.  Deep H$\alpha$ observations have revealed the presence of a vastly extended nebula around the system, which could be the `super-remnant' of many thousands of past eruptions.  With a prediction of the white dwarf reaching the Chandrasekhar mass in less than a mega-year, M31N\,2008-12a has become the leading pre-explosion type Ia supernova candidate.  The 2016 eruption -- to be the twelfth detected eruption -- is expected imminently, and a vast array of follow-up observations are already planned. 
\end{abstract}

\section{Introduction}

Novae are powered by a thermonuclear runaway occurring at the base of an accreted layer at the surface of a white dwarf (WD).  A sub-class of the cataclysmic variables (CVs), novae occur in close binary systems where the WD accretes material from its companion, usually via an accretion disk around the WD.  Classical Novae have been observed in eruption only once, whereas the Recurrent Novae (RNe) have been detected erupting at least twice and have observed recurrence periods up to $\sim100$\,years.  Such short recurrence periods are driven by the combination of a high mass WD and a high mass accretion rate \citep[see][for recent reviews]{2008clno.book.....B,2010AN....331..160B,2012ApJ...746...61D,2014ASPC..490.....W}.  RNe have long been a proposed single-degenerate pathway to Supernovae Type\,Ia (SNe\,Ia), with recent works indicating that the WD mass in nova systems may indeed be growing over time \citep[e.g.][]{2016ApJ...819..168H}.

The remarkable M31N\,2008-12a was first observed in eruption in 2008 \citep{2008Nis}.  Since then, eruptions have been discovered in each year, the most recent in 2015 August --  indicating a mean recurrence period of $347\pm10$\,days \citep{2016arXiv160708082D}.  However, when archival X-ray detections of eruptions from 1992, 1993, and 2001 \citep{2014A&A...563L...8H} are also considered, there is strong evidence that the true recurrence period may be half as long, $174\pm10$\,days \citep{2015A&A...582L...8H}.

In this article, we review a number of the most interesting and unusual features of the M31N\,2008-12a system.  However, we also refer the reader to the extensive published work on this system \citep[see][]{2014A&A...563L...9D,2015A&A...580A..45D,2016arXiv160708082D,2014A&A...563L...8H,2015A&A...582L...8H,2015A&A...580A..46H,2015ApJ...808...52K,0004-637X-830-1-40,2014ApJ...786...61T}.

\section{The white dwarf}

With such a short recurrence period, the inference is that the WD in the M31N\,2008-12a system must be high, and probably the highest mass WD yet found in a CV.  The rapid unveiling of the super-soft source (SSS) X-ray emission after each eruption -- just six days \citep{2016arXiv160708082D} -- points to a low ejected mass \citep[$\sim3\times10^{-8}\,\mathrm{M}_\odot$;][]{2015A&A...582L...8H}; the short duration of the SSS ($t_\mathrm{off}\simeq19$ days) again indicates a high mass WD.  Modelling of the 2013 eruption by \citet{2014ApJ...786...61T} suggested that $M_\mathrm{WD}>1.3\,\mathrm{M}_\odot$ and that the accretion rate must be $>1.7\times10^{-7}\,\mathrm{M}_\odot\,\mathrm{yr}^{-1}$.  Subsequent treatments by \citet{2015ApJ...808...52K}, using data from multiple eruptions and assuming a recurrence period of one year, are consistent with $M_\mathrm{WD}=1.38\,\mathrm{M}_\odot$, $\dot{M}=1.6\times10^{-7}\,\mathrm{M}_\odot\,\mathrm{yr}^{-1}$, and a mass accumulation efficiency of $\eta=0.63$.  Both the  \citet{2014ApJ...786...61T} and \citet{2015ApJ...808...52K} works indicate that the WD in M31N\,2008-12a will reach the Chandrasekhar mass ($\mathrm{M}_\mathrm{Ch}$) in less than 1\,Myr.

With the mass of the WD increasing, the ultimate fate of the system will therefore depend upon the underlying composition of that WD.  An ONe WD will be expected to form a neutron star when $\mathrm{M}_\mathrm{Ch}$ is reached, only a CO WD will explode as a SN\,Ia.  However, in either situation, M31N\,2008-12a is still of vital importance, being so close to $\mathrm{M}_\mathrm{Ch}$ it is either the leading candidate for either an accretion induced collapse or a SN\,Ia.  The recent work of \citet{2016ApJ...819..168H} has shown that a CO WD can grow from its formation mass ($<1.1\,\mathrm{M}_\odot$) to $\mathrm{M}_\mathrm{Ch}$ via a long series of H-flashes (novae) interspersed with He-flashes (He-novae) with little or no tuning of the system parameters or accretion rate.  That work, the discovery of M31N\,2008-12a, and recent RN population work \citep{2014ApJ...788..164P,2015ApJS..216...34S,2016ApJ...817..143W} has significantly strengthened the case for novae contributing to the SN\,Ia progenitor population.

\section{The system}

The high mass accretion rates in RNe are typically provided by either a Roche lobe overflowing sub-giant, as in the U~Scorpii system, or via the stellar wind of a red giant donor \citep[e.g.\ RS~Ophiuchi;][]{2012ApJ...746...61D}.  Using the {\it Hubble Space Telescope} (HST) \citet{2009ApJ...705.1056B} and \citet{2016ApJ...817..143W} have shown that quiescent novae with red giant donors (near-infrared observations) or luminous accretion disks (visible and near-ultraviolet) can be directly imaged at the distance of M31.  Both \citet{2014A&A...563L...9D} and \citet{2014ApJ...786...61T} reported the detection of M31N\,2008-12a at quiescence using archival HST observations.  The SED obtained from those visible and NUV images indicated the presence of an extremely luminous accretion disk, however, the NIR data were too crowded to usefully constrain the donor.

Although the eruptions of M31N\,2008-12a have been followed photometrically from peak back to quiescence \citep{HST2016}, to date we have only been able to follow the eruptions spectroscopically for the first five days post-eruption.  \citet{2016arXiv160708082D} compiled the spectra from the 2012--2015 eruptions and derived the ejecta expansion velocity as a function of time since each eruption.  This was conducted by using the FWHM of the H$\alpha$ emission lines as a proxy of the expansion velocity in the radial direction to the observer.  This analysis clearly indicated that the ejecta velocity decreased as a function of time with $v\propto t^{-1/3}$, consistent with Phase~II shock behaviour \citep[adiabatic expansion, e.g.][]{1985MNRAS.217..205B} as the M31N\,2008-12a ejecta interact with surrounding, pre-existing, and nearby material.  With eruptions occurring regularly in the system, any such circumbinary material must be replenished in between each eruption.

The majority of the spectra taken of M31N\,2008-12a to-date have been at low-resolution, as somewhat necessitated by the faintness and distance of the system.  \citet{2016arXiv160708082D} also presented a co-added spectrum combining all data from the 2012--2015 eruptions.  This deep spectrum indicated the presence of coronal [Fe\,{\sc vii}], [Fe\,{\sc x}], and [Fe\,{\sc xiv}] emission lines as are typically associated with shocks between a nova ejecta and surrounding material.  The deep spectrum also contained an emission band at $\sim6830$\,\AA\ that is consistent with Raman scattering of the O\,{\sc vi} resonant doublet (at 1032/1038\,\AA).  Such Raman scattered features are common in the spectra of symbiotic stars.

Together, the combination of the spectral line development, the coronal emission lines, the Raman emission band, and also the photometric behaviour of each eruption \citep[see][]{2016arXiv160708082D}, provide compelling evidence as to the nature of the mass donor in the system -- all pointing to a red giant, and wind accretion.  This in itself poses an intriguing thought, with the orbital period of novae with red giant donors being of order a year \citep{2012ApJ...746...61D}, M31N\,2008-12a is the only known nova whose orbital period is approximately the same as its recurrence period.  \citet{2016MNRAS.tmp.1538S} have recently indicated that the red giant donor in the RS~Oph system may be tidally locked to its WD and hence the orbits circularised.  Any deviation from circular orbits in the M31N\,2008-12a system however, may lead to an accretion rate, and hence inter eruption interval, that is affected by the orbital phase.

\citet{2016arXiv160708082D} also discussed the accretion disk around the WD, and considered the possibility that the disk survives each eruption.  With a short inter eruption period it is fair to assume that accretion must begin shortly after each eruption.  \citet{2011ApJ...742..113S} argued that the formation timescale of an accretion disk must be at least one orbital period of the system.  Therefore, with a red giant donor, and the recurrence period in M31N\,2008-12a of order the orbital period, it seems unlikely that the disk seen in the HST observations \citep{2014A&A...563L...9D,2014ApJ...786...61T} could re-form in such a short timescale -- therefore the disk must survive.  The accretion disk, and its survival, will be discussed in further detail in \citet{HST2016}.

\section{The `super remnant'}

With one or even two eruptions every year since at least the early 1990s, the search is on to recover some of the missing eruptions.  These efforts led to the discovery of extended H$\alpha$ emission around the nova within archival Steward Observatory \citep{2008ApJ...686.1261C} and Local Group Galaxies Survey data \citep{2007AJ....134.2474M}.  To confirm this emission, a series of deep H$\alpha$ observations were taken by the Liverpool Telescope \citep[LT;][]{2004SPIE.5489..679S}.  Those observations uncovered a shell-like elliptical nebula centred on the position of the nova \citep[see Figure~\ref{fig1} and][]{2015A&A...580A..45D}.  The nebula itself is vast, measuring over 130 pc along the major axis and 90 pc across the minor axis, and is somewhat larger than the majority of Galactic SN remnants.

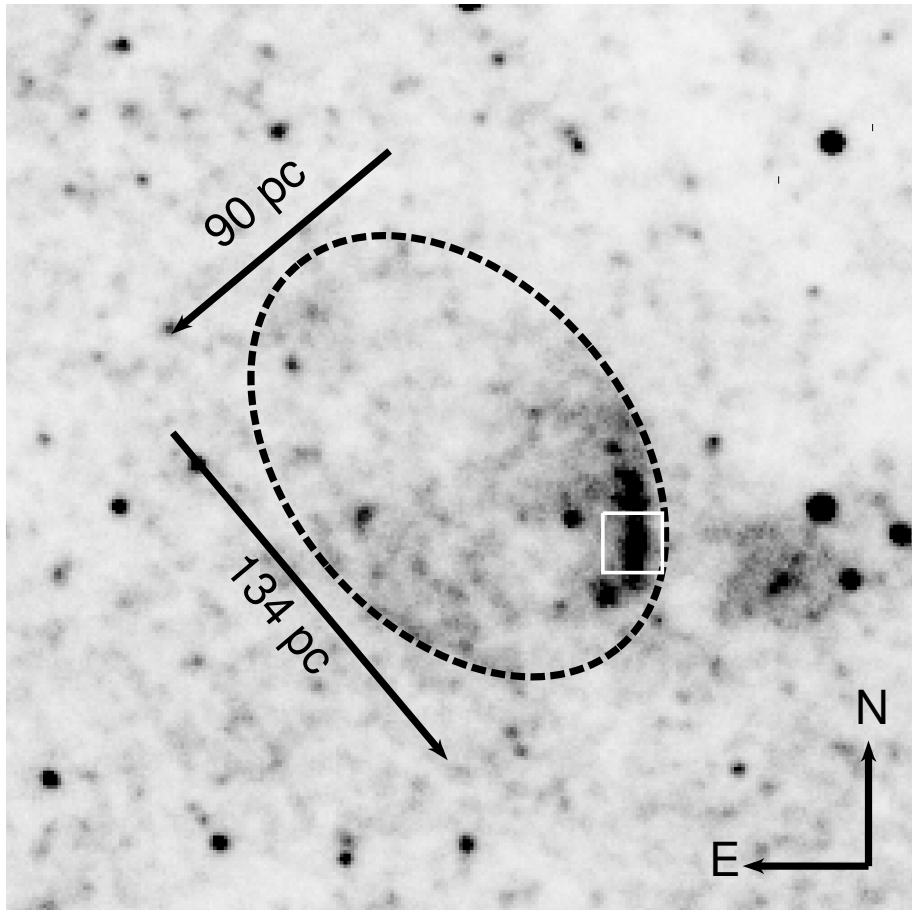
\articlefigure{Darnley_M_Fig1.eps}{fig1}{Liverpool Telescope 68.7\,ks H$\alpha$ image (approximately $1^\prime\!.5\times1^\prime\!.5$) showing the region around M31N\,2008-12a (centre of image).  The dashed ellipse borders the elliptical nebulosity that surrounds the system -- the proposed `super remnant'.  The white box indicates the position of the south-western `knot' (see text for further discussion).}

A spectrum of the south-western `knot' (see Figure~\ref{fig1}) was serendipitously obtained by the long slit of the SPRAT \citep{2014SPIE.9147E..8HP} instrument on the LT during the 2015 eruption \citep{2015A&A...580A..45D}.  That spectrum revealed only three sets of emission lines, H$\alpha$, [N\,{\sc ii}] (6548/6584\,\AA), and [S\,{\sc ii}] (6716/6731\,\AA), no [O\,{\sc iii}] emission was detected in the spectrum nor in narrow-band imaging \citep{2007AJ....134.2474M}.  The [S\,{\sc ii}]/H$\alpha$ ratio of 0.35 is inconclusive as to the nature of the nebula, but the [S\,{\sc ii}] doublet line ratio indicates an electron density that is consistent with the nebula mass being dominated by the ISM swept up from within the shell.

\citet{2015A&A...580A..45D} proposed that the extended nebulosity is the `super-remnant' of many thousands of past eruptions from M31N\,2008-12a.  Hydrodynamic simulations of the formation of such a phenomena are underway, support this hypothesis, and will be reported in a follow-up paper \citep{Hydro}.  Ten orbits of HST Cycle~24 time has been awarded to study the structure and possible formation pathway of the super remnant, with observations due to be taken in 2016 December.

\section{Summary and the future}

Based on the times of the last eleven observed eruptions, \citet{2016arXiv160708082D} predicted that the next observable eruption of M31N\,2008-12a will take place between 2016 August 21 and October 13 ($1\sigma$ range).  A global network of observers has been established to detect this eruption in order to ensure early discovery and the triggering of follow-up observations.  At the time of writing (2016 October 13), this window has now passed without an eruption\footnote{It was hoped that the 2016 eruption would appear in time for this article.}.  When the next eruption is detected, a range of follow-up programmes will be triggered including late time spectroscopy (beyond the 5 day limit so far achieved).  These include observations from Gemini North, Gran Telescopio Canarias (GTC), Hobby Eberly Telescope (HET), and the Large Binocular Telescope (LBT).  In addition to late time spectroscopy of the eruption, they will each obtain deep spectra of the proposed super-remnant, and will attempt spectroscopy at quiescence (a first beyond the Galaxy and Magellanic Clouds).

M31N\,2008-12a is a truly remarkable system, with observed eruptions yearly and a possible six months recurrence period.  The WD in the system is close to the Chandrasekhar mass limit, the accretion rate is high, most likely fed by the stellar wind of a red giant, and the accretion disk appears to survive each eruption.  The system is also surrounded by a huge region of extended nebulosity, which has been proposed to be the super-remnant of many thousands of past eruptions.  The one big outstanding question is the composition of the WD itself, but nevertheless M31N\,2008-12a has become the leading pre-explosion SN\,Ia progenitor candidate.  If confirmed, the super-remnant phenomena may exist around all RNe and could provide an invaluable `sign post' around exploding SNe\,Ia pointing directly to the progenitor type.

The 2016 eruption is due imminently \citep{2016ATel.9415....1D}, we welcome and encourage all observations of this and future eruptions.

\acknowledgements This article is presented on behalf of the `12a' collaboration.  M.J.D. would like to particularly thank Martin Henze for his invaluable help in our on-going campaigns with this remarkable system, and also the LOC and SOC for organising such a wonderful workshop.


\begin{thebibliography}{}
\expandafter\ifx\csname natexlab\endcsname\relax\def\natexlab#1{#1}\fi
\expandafter\ifx\csname url\endcsname\relax
  \def\url#1{\texttt{#1}}\fi
\expandafter\ifx\csname urlprefix\endcsname\relax\def\urlprefix{URL }\fi
\providecommand{\eprint}[2][]{\url{#2}}

\bibitem[{{Bode}(2010)}]{2010AN....331..160B}
{Bode}, M.~F. 2010, Astronomische Nachrichten, 331, 160. \eprint{0911.5254}

\bibitem[{{Bode} et~al.(2009){Bode}, {Darnley}, {Shafter}, {Page}, {Smirnova},
  {Anupama}, \& {Hilton}}]{2009ApJ...705.1056B}
{Bode}, M.~F., {Darnley}, M.~J., {Shafter}, A.~W., et al. 2009, \apj, 705, 1056.
  \eprint{0902.0301}

\bibitem[{{Bode} \& {Evans}(2008)}]{2008clno.book.....B}
{Bode}, M.~F., \& {Evans}, A. (eds.)  2008, {Classical Novae, 2nd Edition},
  vol.~43 of Cambridge Astrophysics Series (Cambridge: Cambridge University
  Press)

\bibitem[{{Bode} \& {Kahn}(1985)}]{1985MNRAS.217..205B}
{Bode}, M.~F., \& {Kahn}, F.~D. 1985, \mnras, 217, 205

\bibitem[{{Coelho} et~al.(2008){Coelho}, {Shafter}, \&
  {Misselt}}]{2008ApJ...686.1261C}
{Coelho}, E.~A., {Shafter}, A.~W., \& {Misselt}, K.~A. 2008, \apj, 686, 1261.
  \eprint{0807.0210}

\bibitem[{{Darnley} \& {Henze}(2016)}]{2016ATel.9415....1D}
{Darnley}, M.~J., \& {Henze}, M. 2016, The Astronomer's Telegram, 9415

\bibitem[{{Darnley} et~al.(2016{\natexlab{a}}){Darnley}, {Henze}, {Bode},
  {Hachisu}, {Hernanz}, {Hornoch}, {Hounsell}, {Kato}, {Ness}, {Osborne},
  {Page}, {Ribeiro}, {Rodriguez-Gil}, {Shafter}, {Shara}, {Steele}, {Williams},
  {Arai}, {Arcavi}, {Barsukova}, {Boumis}, {Chen}, {Fabrika}, {Figueira},
  {Gao}, {Gehrels}, {Godon}, {Goranskij}, {Harman}, {Hartmann}, {Hosseinzadeh},
  {Horst}, {Itagaki}, {Jose}, {Kabashima}, {Kaur}, {Kawai}, {Kennea}, {Kiyota},
  {Kucakova}, {Lau}, {Maehara}, {Naito}, {Nakajima}, {Nishiyama}, {O'Brien},
  {Quimby}, {Sala}, {Sano}, {Sion}, {Valeev}, {Watanabe}, {Watanabe},
  {Williams}, \& {Xu}}]{2016arXiv160708082D}
{Darnley}, M.~J., {Henze}, M., {Bode}, M.~F., et al. 2016{\natexlab{a}}, ApJ, in press, ArXiv e-prints. \eprint{1607.08082}

\bibitem[{{Darnley} et~al.(2015){Darnley}, {Henze}, {Steele}, {Bode},
  {Ribeiro}, {Rodr{\'{\i}}guez-Gil}, {Shafter}, {Williams}, {Baer}, {Hachisu},
  {Hernanz}, {Hornoch}, {Hounsell}, {Kato}, {Kiyota}, {Ku{\v c}{\'a}kov{\'a}},
  {Maehara}, {Ness}, {Piascik}, {Sala}, {Skillen}, {Smith}, \&
  {Wolf}}]{2015A&A...580A..45D}
{Darnley}, M.~J., {Henze}, M., {Steele}, I.~A., et al. 2015, \aap, 580, A45. \eprint{1506.04202}

\bibitem[{{Darnley} et~al.(2016{\natexlab{b}}){Darnley}, {Hounsell}, {Bode},
  {Harman}, {Henze}, {Hornoch}, {Ness}, {Ribeiro}, {Shafter}, {Shara}, \&
  {Williams}}]{HST2016}
{Darnley}, M.~J., {Hounsell}, R., {Williams}, S.~C., et al. 2016{\natexlab{b}}, in preparation

\bibitem[{{Darnley} \& {O'Brien}(2016)}]{Hydro}
{Darnley}, M.~J., \& {O'Brien}, T.~J. 2016, in preparation

\bibitem[{{Darnley} et~al.(2012){Darnley}, {Ribeiro}, {Bode}, {Hounsell}, \&
  {Williams}}]{2012ApJ...746...61D}
{Darnley}, M.~J., {Ribeiro}, V.~A.~R.~M., {Bode}, M.~F., et al. 2012, \apj, 746, 61. \eprint{1112.2589}

\bibitem[{{Darnley} et~al.(2014){Darnley}, {Williams}, {Bode}, {Henze}, {Ness},
  {Shafter}, {Hornoch}, \& {Votruba}}]{2014A&A...563L...9D}
{Darnley}, M.~J., {Williams}, S.~C., {Bode}, M.~F., et al. 2014, \aap, 563, L9.
  \eprint{1401.2905}

\bibitem[{{Henze} et~al.(2015{\natexlab{a}}){Henze}, {Darnley}, {Kabashima},
  {Nishiyama}, {Itagaki}, \& {Gao}}]{2015A&A...582L...8H}
{Henze}, M., {Darnley}, M.~J., {Kabashima}, et al. 2015{\natexlab{a}}, \aap, 582, L8. \eprint{1508.06205}

\bibitem[{{Henze} et~al.(2014){Henze}, {Ness}, {Darnley}, {Bode}, {Williams},
  {Shafter}, {Kato}, \& {Hachisu}}]{2014A&A...563L...8H}
{Henze}, M., {Ness}, J.-U., {Darnley}, M.~J., et al. 2014, \aap, 563, L8.
  \eprint{1401.2904}

\bibitem[{{Henze} et~al.(2015{\natexlab{b}}){Henze}, {Ness}, {Darnley}, {Bode},
  {Williams}, {Shafter}, {Sala}, {Kato}, {Hachisu}, \&
  {Hernanz}}]{2015A&A...580A..46H}
{Henze}, M., {Ness}, J.-U., {Darnley}, M.~J., et al.
  2015{\natexlab{b}}, \aap, 580, A46. \eprint{1504.06237}

\bibitem[{{Hillman} et~al.(2016){Hillman}, {Prialnik}, {Kovetz}, \&
  {Shara}}]{2016ApJ...819..168H}
{Hillman}, Y., {Prialnik}, D., {Kovetz}, A., et al. 2016, \apj, 819,
  168. \eprint{1508.03141}

\bibitem[{{Kato} et~al.(2015){Kato}, {Saio}, \&
  {Hachisu}}]{2015ApJ...808...52K}
{Kato}, M., {Saio}, H., \& {Hachisu}, I. 2015, \apj, 808, 52.
  \eprint{1506.05364}

\bibitem[{{Kato} et~al.(2016)Kato, Saio, Henze, Ness, Osborne, Page, Darnley,
  Bode, Shafter, Hernanz, Gehrels, Kennea, \& Hachisu}]{0004-637X-830-1-40}
Kato, M., Saio, H., Henze, M., et al. 2016, ApJ, 830, 40. \eprint{1607.07985}

\bibitem[{{Massey} et~al.(2007){Massey}, {McNeill}, {Olsen}, {Hodge}, {Blaha},
  {Jacoby}, {Smith}, \& {Strong}}]{2007AJ....134.2474M}
{Massey}, P., {McNeill}, R.~T., {Olsen}, et al. 2007, \aj, 134, 2474.
  \eprint{0709.1267}

\bibitem[{{Nishiyama} \& {Kabashima}(2008)}]{2008Nis}
{Nishiyama}, K., \& {Kabashima}, F. 2008, {CBAT}.
  \url{http://www.cbat.eps.harvard.edu/iau/CBAT\_M31.html\#2008-12a}

\bibitem[{{Pagnotta} \& {Schaefer}(2014)}]{2014ApJ...788..164P}
{Pagnotta}, A., \& {Schaefer}, B.~E. 2014, \apj, 788, 164. \eprint{1405.0246}

\bibitem[{{Piascik} et~al.(2014){Piascik}, {Steele}, {Bates}, {Mottram},
  {Smith}, {Barnsley}, \& {Bolton}}]{2014SPIE.9147E..8HP}
{Piascik}, A.~S., {Steele}, I.~A., {Bates}, S.~D., et al, 2014, SPIE Conference Series, 9147, 8

\bibitem[{{Schaefer} et~al.(2011){Schaefer}, {Pagnotta}, {LaCluyze},
  {Reichart}, {Ivarsen}, {Haislip}, {Nysewander}, {Moore}, {Oksanen},
  {Worters}, {Sefako}, {Mentz}, {Dvorak}, {Gomez}, {Harris}, {Henden}, {Guan
  Tan}, {Templeton}, {Allen}, {Monard}, {Rea}, {Roberts}, {Stein}, {Maehara},
  {Richards}, {Stockdale}, {Krajci}, {Sjoberg}, {McCormick}, {Revnivtsev},
  {Molkov}, {Suleimanov}, {Darnley}, {Bode}, {Handler}, {Lepine}, \&
  {Shara}}]{2011ApJ...742..113S}
{Schaefer}, B.~E., {Pagnotta}, A., {LaCluyze}, A.~P., et al. 2011,
  \apj, 742, 113. \eprint{1108.1214}

\bibitem[{{Shafter} et~al.(2015){Shafter}, {Henze}, {Rector}, {Schweizer},
  {Hornoch}, {Orio}, {Pietsch}, {Darnley}, {Williams}, {Bode}, \&
  {Bryan}}]{2015ApJS..216...34S}
{Shafter}, A.~W., {Henze}, M., {Rector}, T.~A., et al. 2015, \apjs, 216, 34. \eprint{1412.8510}

\bibitem[{{Somero} et~al.(2016){Somero}, {Hakala}, \&
  {Wynn}}]{2016MNRAS.tmp.1538S}
{Somero}, A., {Hakala}, P., \& {Wynn}, G.~A. 2016, \mnras, in press. \eprint{1610.00914}

\bibitem[{{Steele} et~al.(2004){Steele}, {Smith}, {Rees}, {Baker}, {Bates},
  {Bode}, {Bowman}, {Carter}, {Etherton}, {Ford}, {Fraser}, {Gomboc}, {Lett},
  {Mansfield}, {Marchant}, {Medrano-Cerda}, {Mottram}, {Raback}, {Scott},
  {Tomlinson}, \& {Zamanov}}]{2004SPIE.5489..679S}
{Steele}, I.~A., {Smith}, R.~J., {Rees}, P.~C., et al. 2004, in Ground-based
  Telescopes, edited by J.~M. {Oschmann}, Jr., vol. 5489 SPIE Conference Series, 679

\bibitem[{{Tang} et~al.(2014){Tang}, {Bildsten}, {Wolf}, {Li}, {Kong}, {Cao},
  {Cenko}, {De Cia}, {Kasliwal}, {Kulkarni}, {Laher}, {Masci}, {Nugent},
  {Perley}, {Prince}, \& {Surace}}]{2014ApJ...786...61T}
{Tang}, S., {Bildsten}, L., {Wolf}, W.~M., et al. 2014, \apj, 786, 61. \eprint{1401.2426}

\bibitem[{{Williams} et~al.(2016){Williams}, {Darnley}, {Bode}, \&
  {Shafter}}]{2016ApJ...817..143W}
{Williams}, S.~C., {Darnley}, M.~J., {Bode}, M.~F., et al. 2016,
  \apj, 817, 143. \eprint{1512.04088}

\bibitem[{{Woudt} \& {Ribeiro}(2014)}]{2014ASPC..490.....W}
{Woudt}, P.~A., \& {Ribeiro}, V.~A.~R.~M. (eds.) 2014, {Stella Novae: Past and
  Future Decades}, vol. 490 of Astronomical Society of the Pacific Conference
  Series (San Francisco: Astronomical Society of the Pacific)

\end{thebibliography}
\end{document}